\documentclass[prl,twocolumn,showpacs,aps,superscriptaddress]{revtex4}
\usepackage{graphicx,epsfig}
\usepackage{rotate}
\usepackage{color,times}
\usepackage{amsmath,amssymb}
\newcommand{\average}[1]{{\left\langle {#1} \right\rangle}}
\newcommand{\sz}{\scriptsize}

\begin{document}

\title{Diffusion dynamics on multiplex networks}

\author{S. G{\'o}mez}
\affiliation{Departament d'Enginyeria Inform\`{a}tica i Matem\`{a}tiques, Universitat Rovira i Virgili, Tarragona, Spain}

\author{A.  D\'{\i}az-Guilera}
\affiliation{Departament de F\'{\i}sica Fonamental, Universitat de Barcelona, Barcelona, Spain}
\affiliation{Institute for Biocomputation and Physics of Complex Systems (BIFI), Universidad de Zaragoza, 50018 Zaragoza, Spain}

\author{J. G\'omez-Garde\~nes}
\affiliation{Institute for Biocomputation and Physics of Complex Systems (BIFI), Universidad de Zaragoza, 50018 Zaragoza, Spain}
\affiliation{Departamento \ de F\'{\i}sica de la Materia Condensada, Universidad de Zaragoza, 50009 Zaragoza, Spain}

\author{C.J. P\'erez-Vicente}
\affiliation{Departament de F\'{\i}sica Fonamental, Universitat de Barcelona, Barcelona, Spain}

\author{Y. Moreno}
\affiliation{Institute for Biocomputation and Physics of Complex Systems (BIFI), Universidad de Zaragoza, 50018 Zaragoza, Spain}
\affiliation{Departamento \ de F\'{\i}sica Te\'orica, Universidad de Zaragoza, 50009 Zaragoza, Spain}

\author{A. Arenas}
\affiliation{Departament d'Enginyeria Inform\`{a}tica i Matem\`{a}tiques, Universitat Rovira i Virgili, Tarragona, Spain}
\affiliation{Institute for Biocomputation and Physics of Complex Systems (BIFI), Universidad de Zaragoza, 50018 Zaragoza, Spain}

\begin{abstract}
We study the time scales associated to diffusion processes that take place on multiplex networks, i.e.\ on a set of networks linked through interconnected layers. To this end, we propose the construction of a supra-Laplacian matrix, which consists of a dimensional lifting of the Laplacian matrix of each layer of the multiplex network. We use perturbative analysis to reveal analytically the structure of eigenvectors and eigenvalues of the complete network in terms of the spectral properties of the individual layers. The spectrum of the supra-Laplacian allows us to understand the physics of diffusion-like processes on top of multiplex networks.
\end{abstract}


\maketitle

Modern theory of complex networks is facing new challenges that arise from the necessity of understanding properly the dynamical evolution of real systems. One of such open problems concerns the topological and dynamical characterization of systems made up by two or more interconnected networks. The standard approach in network modeling assumes that every edge (link) is of the same type and consequently considered at the same temporal and topological scale \cite{newman2010}.
\begin{figure}[b]
\begin{center}
\includegraphics[width=0.45\textwidth]{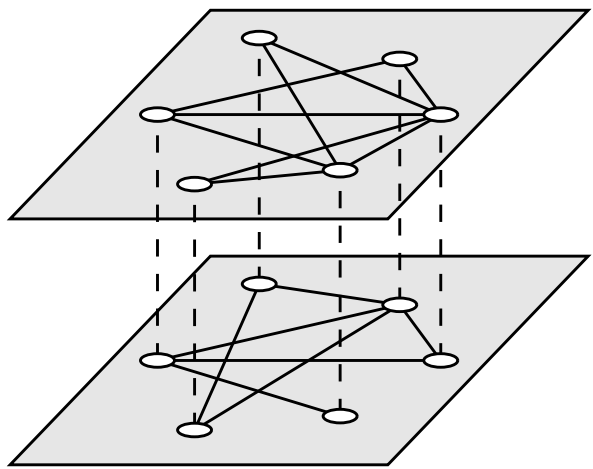}
\end{center}
\caption{Example of multiplex network with $M=2$ layers. Nodes are the same in both layers. The connectivity at each layer is independent of each other, the connectivity inter-layer is from each node to itself (dashed links).}
\label{fig:multiplex}
\end{figure}This is clearly an abstraction of any real topological structure and represents either instantaneous or aggregated interactions over a certain time window. Therefore, to understand the intricate variability of real complex systems, where many different time scales and structural patterns coexist we need a new scenario, a new level of description \cite{mucha2010}.

A natural extension which allows to overcome previous drawbacks is to describe a multilevel system as a set of
coupled layered networks ({\em multiplex} network) where each layer could have very particular features different from the rest, and in this way, define a richer structure of interactions \cite{koreans12}. Multiplex networks are thus structured multilevel graphs in which interconnections between layers determine how a given node in a layer and its counterpart in another layer are linked and influence each other. Thus, they are essentially different from simple graphs with colored edges, multi graphs or hypergraphs and provide a mathematical ground for the analysis of many social networks (e.g.\ Facebook, Twitter, etc) and of several biological systems $-$ for instance, in biochemical networks, many different signaling channels do actually work in parallel, giving raise to what is called multitasking, which can be modeled through a network of interconnected layers \cite{cozzo}. Although some works have recently focused on the description and analysis of interconnected networks \cite{par10,buldy10,hu11,gao11,gao12}, theoretically grounded results about general dynamical processes running on them are yet to come.

In this Letter we focus on a particular setup of multilevel networks in which nodes are conserved through the different layers of the multiplex (see Fig.~\ref{fig:multiplex}). The current study analyzes a diffusion process that takes place at the whole system level, i.e.\ within and across layers. This setup could account, for instance, for a diffusion dynamics taking place on top of a social network of contacts. Admittedly, the latter is a network of networks, i.e.\ the aggregate of many different social circles or subnetworks, each having its own temporal or structural patterns (for example, think of our online activity which includes different social networking sites such as Facebook, Twitter, etc.). The same applies to multimodal transportation networks \cite{barthelemy11}, on top of which individuals ``diffuse'' within and between different layers (e.g.\ bus, subway, etc.). Let us remark, however, that our interest here is not to solve a specific real problem but to illustrate the analysis of diffusion processes on top of these structures.

We propose a mathematical setting that allows to scrutinize the emergent diffusion time scales in multiplex networks.
We concentrate on diffusive processes, as they constitute a good approximation for different types of dynamical processes (e.g.\ synchronization and other nonlinear processes amenable of linearization \cite{adkmz08}) whose dynamical properties can be captured by the behavior of the eigenvalues of the Laplacian matrix. For instance, the time needed to synchronize phase oscillators in a network is related to the second smallest eigenvalue of the Laplacian, $\lambda_2$ \cite{ad07}, and the stability of the synchronized state is determined by the eigenratio $\lambda_N/\lambda_2$ \cite{bp02}. The spectral analysis of complex networks constitutes then still a promising area of research \cite{jalan11, estrada12}. Following a perturbative analysis of the spectra \cite{vanmi}, our results allow to get new physical insight about diffusion processes through the analytical determination of the asymptotic behavior of the eigenvalues of the Laplacian of the multiplex (supra-Laplacian) when the diffusive coupling between layers is either small or large. Our findings prove that the emergent physical behavior of the diffusion process when considering coupled layered networks is far from trivial, in some cases (specified below) the coupling of networks shows a super-diffusive behavior meaning that diffusive processes in the multiplex are faster than in any of the networks that form it separately.


Let us consider a set up in which the diffusive dynamics is linearly coupled within nodes in each layer $K$, through a diffusion constant $D_K$, and among nodes in different layers $K$ and $L$, in this case with a diffusion constant $D_{KL}$. The network at each layer is assumed to be connected and undirected, but it can be weighted. The state of each of the $N$ nodes is represented as a vector indexed by layers $x_i^K(t)$ where the subscript stands for the node and the superscript for the layer. The equations describing the dynamical evolution of the states of the nodes, considering a multiplex of the $M$ layers, are:
\begin{equation}
\frac{dx_i^K}{dt}= D_K \sum_{j=1}^{N} w_{ij}^{K}(x_j^K-x_i^K) + \sum_{L=1}^M D_{KL} (x_i^L-x_i^K)\;,
\label{ori}
\end{equation}
\noindent where $w_{ij}^K$ denotes the weight matrix at layer $K$ ($w_{ij}^{K} = 0$ means that there is no link between nodes $i$ and $j$ in layer $K$). This set of equations can be dimensionally lifted to a space of $N\times M$ dimensions. To have a more clear picture of our formalism we will consider, without loss of generalization, the most simple case of two layers $M=2$. First, we define a column vector state of $2N$ elements, $(x_1^1 \cdots x_N^1 | x_1^2 \ldots, x_N^2) = ({\mathbf x}^1 | {\mathbf x}^2 ) = {\mathbf x}$. Then Eq.~(\ref{ori}) can be written in matrix form, where the interaction matrix has a block structure that conforms an object we call supra-Laplacian ${\cal L}$, with the same properties that any zero-sum rows Laplacian has:
\begin{equation}
{\cal L}
=
\left(
\begin{array}{c|c}
D_1 L_1 + D_x  I & -D_x I \\
\hline
-D_x I  &  D_2 L_2 + D_x  I\\
\end{array}
\right)\;,
\end{equation}
\noindent where $L_1$ and $L_2$ are the respective Laplacians of each layer, and $I$ is the identity matrix. Here we have replaced $D_{12}$ by $D_x$ to emphasize the role of the diffusion process among the same node at different layers. The Laplacian matrix of each layer $K$ is just $L_K=S_K-W_K$, where $W_K$ is the weights matrix at layer $K$, and $S_K$ a diagonal matrix containing the strength of each node $i$ at layer $K$, $(S_K)_{ii}=s_i^K=\sum_j w_{ij}^{K}$.
Note that the diagonal block structure of the supra-Laplacian reflects the interaction within layers and the off-diagonal blocks the connectivity between layers.

The dynamical properties of the system can then be cast in terms of the eigenvalues of this matrix. Eq.~(\ref{ori}) can be written as $\dot{{\mathbf x}} = -{\cal L} {\mathbf x}$ and, given that  $\cal L$ is symmetric, its solution in terms of normal modes is $\phi_i (t) = \phi_i (0) e^{-\lambda_{i} t}$, where $\lambda_{i}$ are the eigenvalues of $\cal L$, see e.g.\ \cite{adp06a,adp06b}. The diffusion time scale $\tau$ of the multiplex is controlled by the smallest non-zero eigenvalue of ${\cal L}$. Specifically, $\tau=1/\lambda_2$. To get a physical insight on these eigenvalues as a function of the different diffusion coefficients within layers ($D_1$ and $D_2$) and between layers ($D_x$), we propose to analyze the whole system using perturbation theory. To simplify the notation, we choose the diffusion coefficients $D_1=D_2=1$ fixing then the relative time scale of the problem.

Let us consider the decomposition ${\cal L} = {\cal L}_{0} + \cal D$, where ${\cal L}_{0}$ is the block diagonal matrix corresponding to the Laplacians of every layer, with zeros in the off-diagonal blocks, and ${\cal D}$ is formed by the rest of the elements. In matrix form it reads:
\begin{equation}
{\cal L}
=  {\cal L}_{0} + {\cal D} =
\left(
\begin{array}{c|c}
L_1 & 0 \\
\hline
0  &  L_2\\
\end{array}
\right)
+ { D}_{x}
\left(
\begin{array}{r|r}
I & -I \\
\hline
-I &  I\\
\end{array}
\right)\;.
\end{equation}

Let us start the discussion by considering $D_x=0$. Then, the eigenvalues of ${\cal L}$ are the set formed by the union of the eigenvalues corresponding to the Laplacians of each layer $L_1$ and $L_2$. The eigenvalues are $0=\lambda_1^1< \lambda_2^1\le \ldots \lambda_N^1$ and $0=\lambda_1^2< \lambda_2^2\le \ldots \lambda_N^2$, respectively, while the eigenvalues of the supra-Laplacian matrix are $0=\lambda_1=\lambda_2< \lambda_3\le \ldots \le\lambda_{2n}$, being $\lambda_3 = min(\lambda_2^1,\lambda_2^2)$. It is interesting to note that to analyze the eigenvector space it is convenient to move to a new basis where the space corresponding to $\lambda_1 = \lambda_2 =0$ is spanned by vectors $(1 \cdots 1 | 1 \cdots 1)$ and $(1 \cdots 1 | -1 \cdots -1)$ instead of the canonical $(1 \cdots 1 | 0 \cdots 0)$ and $(0 \cdots 0 | 1 \cdots 1)$.

Now let us consider that the diffusion between layers is
different from zero, $D_{x} \ne 0$.
In this case, the supra-Laplacian will have the trivial eigenvalue $\lambda_1 = 0$ with corresponding eigenvector $(1 \cdots 1 | 1 \cdots 1)$, and a non-trivial eigenvalue $\lambda=2 D_x$ that corresponds exactly to the eigenvector $(1 \cdots 1 | -1 \cdots -1)$, since
\begin{equation}
{\cal L}
\left(
\begin{array}{r}
{\mathbf 1}\\
\hline
-{\mathbf 1}\\
\end{array}
\right)
=
\left(
\begin{array}{r}
{\mathbf 0}\\
\hline
{\mathbf 0}\\
\end{array}
\right)
+
2 D_{x}
\left(
\begin{array}{r}
{\mathbf 1}\\
\hline
-{\mathbf 1}\\
\end{array}
\right)\;.
\end{equation}
Note that this eigenvalue always exists, but it will be the smallest non-zero one only when $D_x$ is very small, as compared to $D_1$ and $D_2$.

Next, we focus our attention on the opposite limit, a very large diffusion coefficient \footnote{Strictly speaking if $D_x=\infty$ the model (\ref{ori}) does not hold and the whole multiplex can be understood as a single projected network with a unique state per node} between layers $D_x \gg 1$. Defining $D_x=1/\epsilon$, we can write
\begin{equation}
{\cal L}
=
D_x
\left[
\left(
\begin{array}{r|r}
I & -I \\
\hline
-I &  I\\
\end{array}
\right)
+
\epsilon
\left(
\begin{array}{c|c}
L_1 & 0 \\
\hline
0  &  L_2\\
\end{array}
\right)
\right] = D_x{\tilde{\cal L}}.
\label{cinco}
\end{equation}
The spectrum of ${\tilde{\cal L}}$ is considered here a perturbation of that at $\epsilon=0$. It is worth recalling that, for $\epsilon=0$, the spectrum corresponds to that of the coupling matrix
\begin{equation}
\left(
\begin{array}{r|r}
I & -I \\
\hline
-I &  I\\
\end{array}
\right),
\end{equation}
which consists of two eigenvalues
$(\tilde{\lambda}_1 =0$ and $\tilde{\lambda}_2=2$) both $N$-degenerate and spanned by eigenvectors of the form $({\mathbf u}|{\mathbf u})$ and $({\mathbf u}|-{\mathbf u})$, i.e.\ vectors having identical or opposite values in the $i^{th}$ and $(i+N)^{th}$ components, respectively. Thus, in the limit $D_x\rightarrow \infty$, the set of eigenvalues of ${\cal L}$ will split in two groups, one showing a linear divergent behavior $\lambda \approx 2D_x$ for the sub-space $({\mathbf u}|-{\mathbf u})$, and another having a finite value $\lambda$ as the result of the undetermined limit ($0\cdot \infty$) in Eq.~(\ref{cinco}) for the sub-space $({\mathbf u}|{\mathbf u})$.

Now, we use the common ansatz in perturbation theory and propose a perturbed solution in terms of both eigenvalues and eigenvectors:
\begin{eqnarray}
\lambda_i &=& \lambda_i^{(0)} + \epsilon \lambda_i^{(1)} + O(\epsilon^{2})\;,\\ \nonumber
{\bf v}_i &=& {\bf v}_i^{(0)} + \epsilon {\bf v}_i^{(1)} + O(\epsilon^{2})\;,
\end{eqnarray}
\noindent where the super-indices within parentheses represent the order of the perturbation \cite{marcus01,chauhan09}.
Given that a set of eigenvalues of ${\cal L}$ will diverge linearly as $2D_x$, we concentrate in proposing perturbations for the finite solutions. These correspond to consider the following perturbation of the eigenspectrum of ${\tilde{\cal L}}$:
\begin{eqnarray}
\tilde{\lambda} &=& 0 + \epsilon \tilde{\lambda}'\;, \nonumber \\
{\bf v} &=& \left(
\begin{array}{r}
{\bf u}\\
\hline
{\bf u}\\
\end{array}
\right)
+ \epsilon
\left(
\begin{array}{r}
{\bf u}_{1}'\\
\hline
{\bf u}_{2}'\\
\end{array}
\right).
\label{pert}
\end{eqnarray}
Expanding to $O(\epsilon)$ the eigenvalue problem $\tilde{\cal L} {\mathbf v} =  \tilde{\lambda} {\mathbf v}$ we obtain:
\begin{equation}
\epsilon \left(
\begin{array}{r}
({\bf u}_{1}' - {\bf u}_{2}') + L_1 {\bf u}\\
\hline
({\bf u}_{2}' - {\bf u}_{1}') + L_2 {\bf u}\\
\end{array}
\right)=\epsilon \tilde{\lambda}' \left(
\begin{array}{r}
{\bf u}\\
\hline
{\bf u}\\
\end{array}
\right) + O(\epsilon^{2})\;.
\label{eq8}
\end{equation}
Matching each of the components in Eq.~(\ref{eq8}) we get:
\begin{eqnarray}
L_1 {\bf u} + ({\bf u}_{1}' - {\bf u}_{2}') &= \tilde{\lambda}' {\bf u} \nonumber\;,\\
L_2 {\bf u} + ({\bf u}_{2}' - {\bf u}_{1}') &= \tilde{\lambda}' {\bf u}\;,\label{dos}
\end{eqnarray}
that, after adding and subtracting Eqs.~(\ref{dos}), transform into:
\begin{eqnarray}
(L_1 + L_2) {\bf u} &=& 2\tilde{\lambda}' {\bf u} \nonumber\\
(L_1 - L_2) {\bf u} &=&  2({\bf u}_{1}' - {\bf u}_{2}') \label{resta}
\end{eqnarray}

From the system of Eqs.~(\ref{resta}) it is revealed that ${\bf u}$ is an eigenvector of the network formed by the superposition of both layers' laplacians, and that the eigenvalue of ${\cal L}$, at first order in the expansion, is
\begin{equation}
\lambda = \tilde{\lambda}' = \frac{\lambda_s}{2},
\end{equation}
being $\lambda_s$ the eigenvalue of the superposition $(L_1+L_2)$ corresponding to the eigenvector ${\bf u}$. Moreover, given that the vector perturbation in Eq.~({\ref{pert}}) must be orthogonal  $({\mathbf u}|{\mathbf u}) \perp  ({\mathbf u}_1 '|{\mathbf u}_2 ')$, we can also find the eigenvector of the superposition $(L_1+L_2)$ such that ${\mathbf u}_2 '= -{\mathbf u}_1 ' \equiv -{\mathbf u}'$, then
\begin{equation}
{\bf u}'= \frac{1}{4} (L_2 - L_1) {\bf u}.
\end{equation}

Summarizing, the eigenvectors with finite (i.e.\ non divergent) eigenvalues of the supra-Laplacian ${\cal L}$ for a large value of the diffusion coefficient $D_x=1/\epsilon$ between layers are
\begin{equation}
{\bf v}'= \left(\frac{{\bf u}+\epsilon{\bf u}'}{{\bf u}-\epsilon{\bf u}'}\right) \hspace{0.5cm} \mbox{with eigenvalue} \hspace{0.5cm}\frac{\lambda_s}{2},
\end{equation}
\noindent being $\mathbf u$ and $\lambda_s$ the eigenvectors and corresponding eigenvalues of the superposition $(L_1+L_2)$.

The physical insight obtained is the following, for low values of the diffusion coefficient between layers, the diffusion time scale of the global system is controlled by the inverse of  $2D_x$. This asymptotic result is valid until the order of $D_x$ is similar to those of $D_1$ and $D_2$. For large values of $D_x$ the eigenspectrum splits into a set of values that diverges as $2D_x$, and a set of finite values, associated to the superposition of the layers. The minimal eigenvalue different from zero turns out to be half the eigenvalue corresponding to the superposition of both layers $\lambda_s /2$.

A comparison between the diffusion time scale of the independent layers and the whole multiplex is possible using known bounds about the eigenvalues of the Laplacians \cite{m91}. The time scale associated to the multiplex for $D_x\ll1$ is $\tau=\frac{1}{2D_x}$ that means that the cross-diffusion between layers is the limiting value of the diffusion spreading. On the other hand, the time scale associated to the multiplex for $D_x\gg1$ is $\tau \approx 2/\lambda_s$. This latter case is far less trivial than the previous one. Using the bounds in  \cite{m91} we deduce the following result:
\begin{equation}
\frac{\lambda_s}{2}\geq \frac{\lambda_2^1 + \lambda_2^2}{2}\geq\min{(\lambda_2^1,\lambda_2^2)}\;.
\label{suma}
\end{equation}
\noindent The above inequality implies that the diffusion in the multiplex will be faster than the diffusion in the slowest layer. Thus, as a consequence of the multiplex structure, at least one layer (the one with the largest diffusion time scale) has its diffusion speeded up. The emergence of a super-diffussion, i.e.\ the fact that the time scale of the multiplex is faster than that of both layers acting separately is, in general, not guaranteed and depends on the specific structures coupled together. Furthermore, the following inequality also holds \cite{m91}:
\begin{equation}
\frac{\lambda_s}{2} \leq \frac{2N}{2N-1} \min_i{\left(\frac{s_i^1+s_i^2}{2}+D_x \right)}\;,
\label{mini}
\end{equation}
\noindent being $s_i^K$ the strength of node $i$ at layer $K$.

Finally, it is worth noticing that although the previous analysis assumes that the networks within layers are connected, we have also analyzed the case in which this hypothesis is relaxed. Imagine for example two layers such as one layer has two disconnected components. In this situation, the results hold in the limit $D_x\gg1$, and in the limit $D_x \ll 1$ the lowest (different from zero) eigenvalue scales as $\alpha D_x$, with $0<\alpha\leq 2$ although the perturbed eigenvector is far more intricate.

\begin{figure}[t]
\begin{center}
\begin{tabular}{c}
\includegraphics[width=0.45\textwidth]{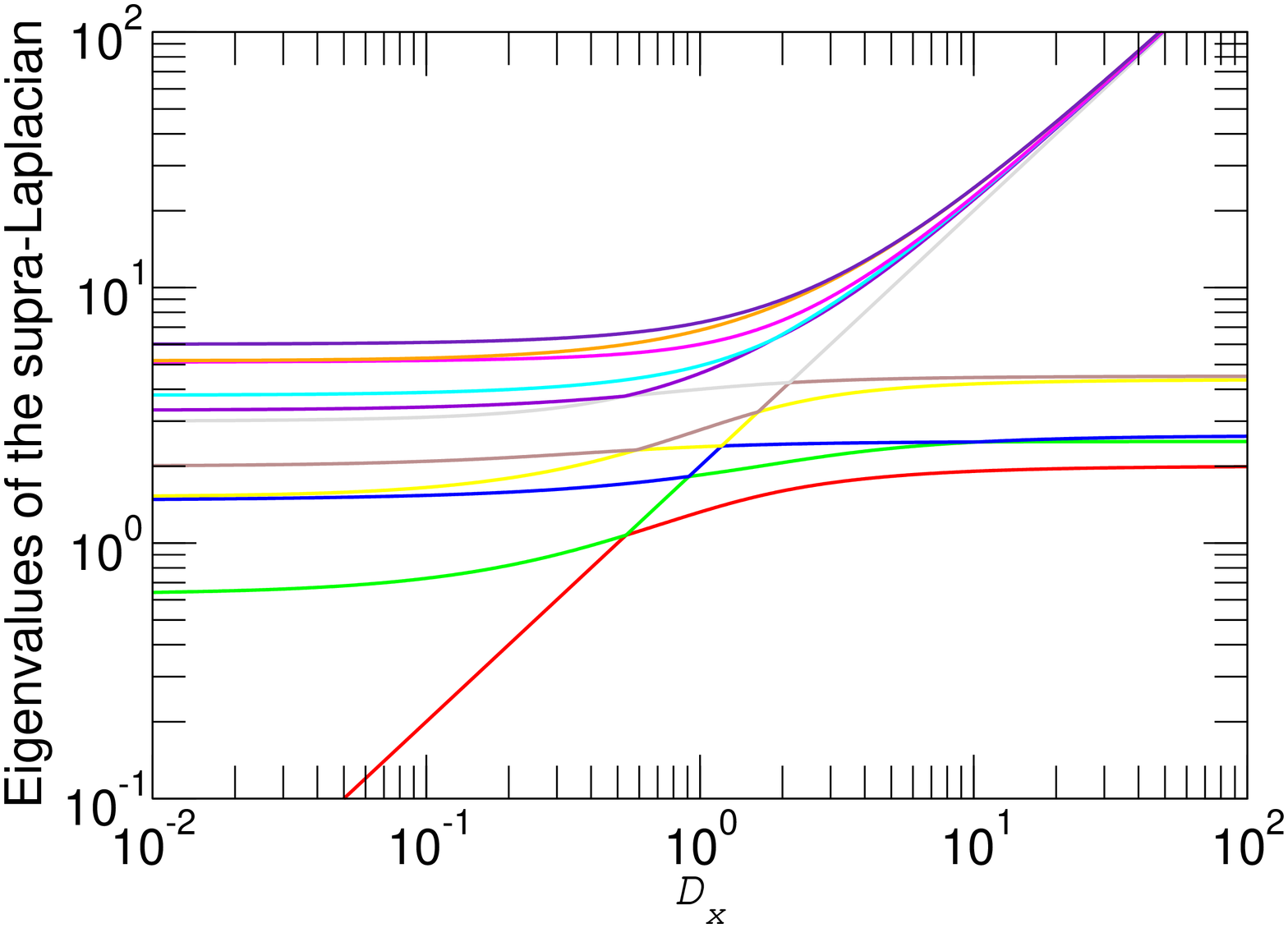} \\ \\
\includegraphics[width=0.45\textwidth,clip=true]{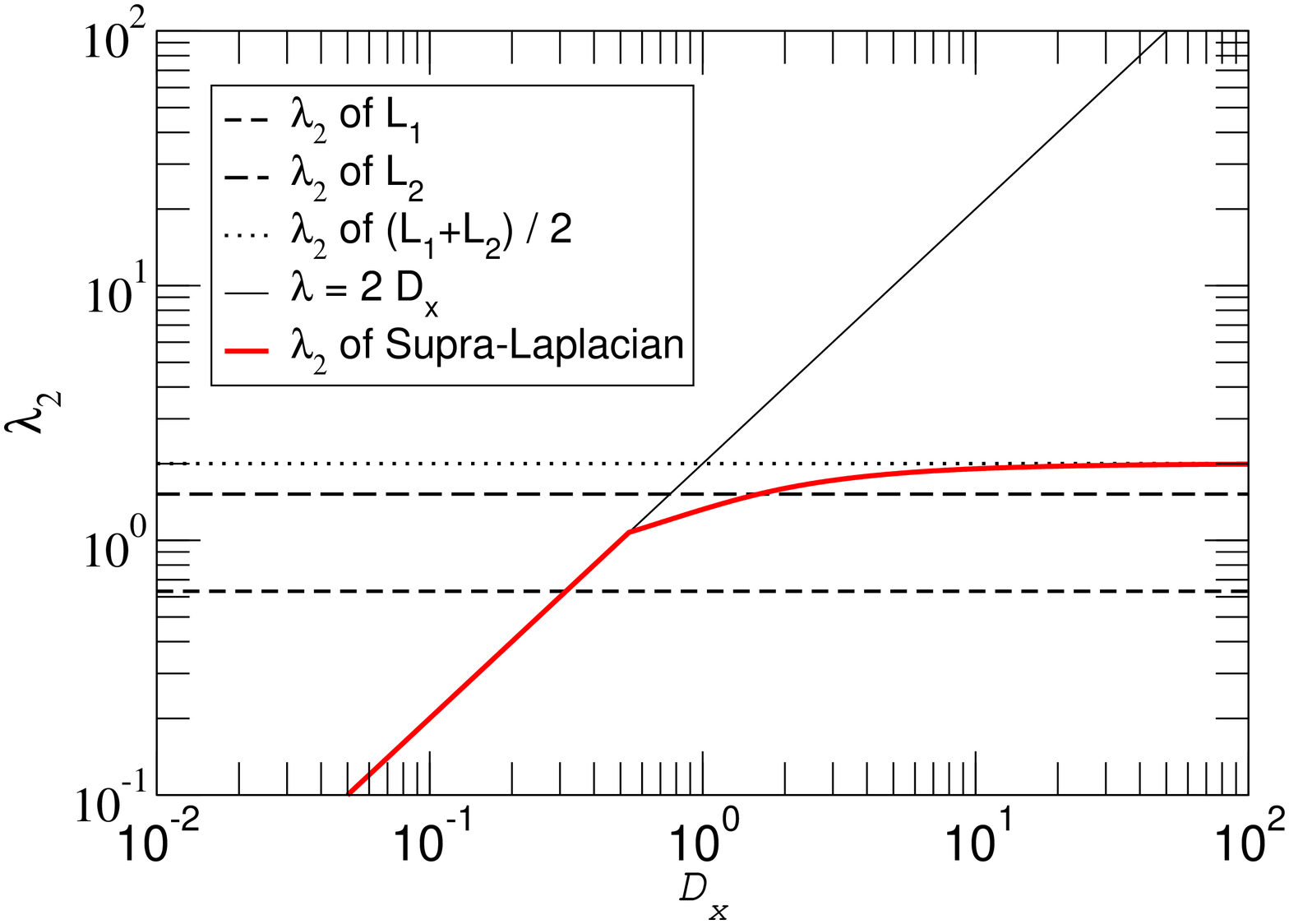}
\end{tabular}
\end{center}
\caption{(color online) Evolution of the eigenspectra of the toy model presented in Fig.~\ref{fig:multiplex} as a function of the coupling $D_x$ (top) for $D_1=D_2=1$, and comparison between the second smallest eigenvalues $\lambda_2$ of the different Laplacians (bottom).}
\label{fig:rand6}
\end{figure}

To illustrate our results, we have computed the evolution of the eigenvalues of the supra-Laplacian for the example represented in Fig.~\ref{fig:multiplex}, which corresponds to two random networks of $N=6$ nodes. In Fig.~\ref{fig:rand6} (top) we plot the eigenvalues as a function of the diffusion coefficient $D_x$.
We observe the splitting of the eigenvalues into two groups, divergent and finite values, as predicted. Fig.~\ref{fig:rand6} (bottom) shows the theoretical estimates for $\lambda_2$ in the asymptotic limits $D_x\ll1$ and $D_x \gg1$. Note that, except for the intermediate zone ($D_x \approx 1$), where the analysis does not hold, the agreement is excellent. In this panel we have represented, as indicated in the legend, the eigenvalues of each layer, the eigenvalue of the superposition of both layers and the line corresponding to $2D_x$  as well as the eigenvalue of the supra-Laplacian. The results undoubtedly confirm that both theoretical limits (small and large $D_x$) are correctly identified by the analytical derivations. Note that the model allows switch on and off the consideration of isolated layers or the whole multiplex, simply by putting $D_x=0$. For the example exposed, we observe a super-diffusion process for the whole multiplex, that means that the time scale associated to the whole multiplex network is smaller than that of layer 1 and layer 2 if they were considered independently, i.e.\ $\tau<\tau_1<\tau_2$. Other examples comparing multiplex networks with 1000 nodes per layer, with different standard topologies, including clustered networks, are presented in the Supplemental Material accompanying this letter, all them showing perfect agreement with the developed analysis.

%

\medskip
In conclusion, we have developed a formalism to unveil the time scales of diffusive processes on multiplex networks. The approach has been specifically presented for a two-layer multiplex, in a particular set up in which nodes are preserved through layers. We obtained analytical results in the two asymptotic limits of small and large diffusion coefficients between layers. The findings show that the multiplex structure is able to speed up the less diffusive of the layers. In principle, it could also give rise to a super-diffusion process thus enhancing the diffusion of both layers. This striking result appears when one considers that the diffusion between the layers of the multiplex is faster than that occurring within each of the layers. Thus, it paves the way to the analysis of super-diffusion processes in real multiplex scenarios such as multimodal transportation systems. On more general grounds, given the wide applicability of the properties of the Laplacian to address many dynamical properties of networked systems, our results constitute a first step towards a better understanding of linear and nonlinear processes on top of multiplex structures.

\acknowledgments This work has been partially supported by the
Spanish DGICYT Grants FIS2009-13364-C02-01, FIS2009-13730-C02-02, FIS2008-01240, FIS2011-25167, MTM2009-13848, FET projects PLEXMATH (317614), MULTIPLEX (317532) and LASAGNE (318132), and the Generalitat de Catalunya 2009-SGR-838. J.G.-G. is supported by the MICINN through the Ram\'on y Cajal Program. A.A. acknowledges the ICREA Academia Award.


\appendix
\section{Supplemental material}

\begin{center}
\includegraphics[width=0.45\textwidth,clip=true]{sf1000g25+sf1000g30-diffusion-lambda2.eps}
\end{center}
Comparison between the second smallest eigenvalues $\lambda_2$ of the different
laplacians for a multiplex network consisting of two layers with 1000 nodes in each layer.
The first contains a scale-free network with degree distribution $P(k)\sim k^{-2.5}$, and
the second layer a scale-free network with degree distribution $P(k)\sim k^{-3}$.

\begin{center}
\includegraphics[width=0.45\textwidth,clip=true]{sf1000g25+er1000k8-diffusion-lambda2.eps}
\end{center}
Comparison between the second smallest eigenvalues $\lambda_2$ of the different
laplacians for a multiplex network consisting of two layers with 1000 nodes in each layer.
The first layer contains a scale-free network with degree distribution $P(k)\sim k^{-2.5}$, and
the second layer contains a random Erd\"os-R\'enyi network with average degree $\average{k}=8$.

\begin{center}
\includegraphics[width=0.45\textwidth,clip=true]{sf1000g25+sw1000r03-diffusion-lambda2.eps}
\end{center}
Comparison between the second smallest eigenvalues $\lambda_2$ of the different
laplacians for a multiplex network consisting of two layers with 1000 nodes in each layer.
The first contains a scale-free network with degree distribution $P(k)\sim k^{-2.5}$, and
the second layer a small-world network with average degree $\average{k}=8$ and a replacement
probability $r=0.3$.

\newpage

\begin{center}
\includegraphics[width=0.45\textwidth,clip=true]{sf1000g25+grid40x25k8-diffusion-lambda2.eps}
\end{center}
Comparison between the second smallest eigenvalues $\lambda_2$ of the different
laplacians for a multiplex network consisting of two layers with 1000 nodes in each layer.
The first contains a scale-free network with degree distribution $P(k)\sim k^{-2.5}$, and
the second layer a $40\times 25$ lattice with eight neighbors per node and periodic
boundary conditions.

\begin{center}
\includegraphics[width=0.45\textwidth,clip=true]{sf1000g25+sf1000g25+20pc-diffusion-lambda2.eps}
\end{center}
Comparison between the second smallest eigenvalues $\lambda_2$ of the different
laplacians for a multiplex network consisting of two layers with 1000 nodes in each layer.
The first contains a scale-free network with degree distribution $P(k)\sim k^{-2.5}$, and
the second layer has been obtained from a copy of the first layer network with 400 extra
random links. Here we observe the absence of super-diffusion. This is a consequence of
the semi-superposition $(W_1+W_2)/2$ being a (weighted) spanning graph of the network
in the second layer $W_2$, thus according to Corollary~3.4 in [20] we have
$\lambda_2((L_1+L_2)/2) \leq \lambda_2(L_2)$.
\\
\\
\\
\\

\begin{center}
\includegraphics[width=0.45\textwidth,clip=true]{ng1000i01e01+ng1000i02e001-diffusion-lambda2.eps}
\end{center}
Comparison between the second smallest eigenvalues $\lambda_2$ of the different
laplacians for a multiplex network consisting of two layers with 1000 nodes in each layer.
The first contains a network structured in 4 communities of 250 nodes each, with average
internal and external degrees
$\langle k^{\mbox{\sz int}}\rangle=105$ and $\langle k^{\mbox{\sz ext}}\rangle=105$ respectively,
and the second layer is similar but with
$\langle k^{\mbox{\sz int}}\rangle=200$ and $\langle k^{\mbox{\sz ext}}\rangle=10$.
The average clustering coefficients at each layer are 0.2336 and 0.7307 respectively,
and the communities in both layers match.

\begin{center}
\includegraphics[width=0.45\textwidth,clip=true]{ng1000_4i105e105+ng1000_17i50e8-diffusion-lambda2.eps}
\end{center}
Comparison between the second smallest eigenvalues $\lambda_2$ of the different
laplacians for a multiplex network consisting of two layers with 1000 nodes in each layer.
The first contains a network structured in 4 communities of 250 nodes each, with average
internal and external degrees
$\langle k^{\mbox{\sz int}}\rangle=105$ and $\langle k^{\mbox{\sz ext}}\rangle=105$ respectively,
and the second layer is similar but with 17 communities,
$\langle k^{\mbox{\sz int}}\rangle=50$ and $\langle k^{\mbox{\sz ext}}\rangle=8$.
The average clustering coefficients at each layer are 0.2336 and 0.6541 respectively.

\end{document}